# Linear-circular proton accelerator with energy 0.8 GeV based on a rectangular cavity with $E_{110}$ wave

S. N. Dolya

*Joint Institute for Nuclear Research, Joliot – Curie Street 6, Dubna, Russia, 141980*

## Abstract

We consider acceleration of protons in a cavity with such dimensions as length l = 10 m, width b = 20 m, height a = 1.01 m, located between two bending ($180^0$) magnets with isochronous magnetic field. The cavity is loaded with rows of drift plates (tubes), being excited by a wave of $E_{110}$ type at the frequency $f_0$ = 148.5 MHz, excitation power $P_1$ = 5.65 MW. Additionally, the cavity is fed with the power $P_2$ = 8 MW, transmitted into a beam of protons with such parameters as the final energy 0.8 GeV, pulse beam current I = 10 mA, pulse width 25 microseconds, repetition rate 10 Hz. We discuss the possibility of creating a pulsed neutron source with the intensity $I_n = 3 * 10^{14}$ neutrons /s on the basis of such an accelerator.

Nowadays, there is still a pretty strong "neutron" hunger in the studies on condensed-matter physics. This means that currently available neutron sources are obviously fewer than tasks that require neutrons for solving.

**Pulsed neutron sources**

The operating principle of a neutron source based on the spallation reaction is well known [1]. When a high-energy proton is accelerated into a heavy target, a number of spallation particles, including neutrons, are produced. For every proton striking the nucleus, 20 to 30 neutrons are expelled. Meson production limits spallation efficiency above 140 MeV. At the 1 GeV proton energy level, the Spallation Neutron Source will require 30 MeV per neutron produced. Neutron scattering is used by a variety of scientific disciplines to study the arrangement, motion, and interaction of atoms in materials. It is important because it provides valuable information that often cannot be obtained using other techniques, such as optical spectroscopies, electron microscopy, and x-ray diffraction. Scientists need all these techniques to provide the maximum amount of information on materials.

The essence of the proposal is to create a pulsed neutron source based on a proton accelerator with the following parameters: the proton energy $W_{fin}$ = 0.8 GeV, pulsed proton beam current $I_p$ = 10 mA, pulse duration $\tau_p$ = 25 μs, repetition rate F = 10 Hz.

The logic of constructing pulsed neutron sources is as follows. It is convenient to measure the speed of the scattered neutrons by the time of flight of the neutrons from the target to the detector. The speed of the thermal neutron $V_n = 2.2$ km/s, and at a reasonable length of the channel $l_{ch} = 10$ m the neutron will travel this distance in a time $\tau_1 = 5$ ms. The resolution 1% by the time of flight is considered "good", which means that duration of the neutron pulse should be shorter than $\tau_2 < 50$ microseconds.

The repetition rate of the accelerator cannot be too big – in order to understand the physics processes correctly, fast neutrons from the next flash are required to reach the target after slow (thermal) neutrons from the previous operation. Since the time of flight for thermal neutrons in a channel of length $l_{ch} = 10$ m is about 5 milliseconds, the time between the subsequent outbreaks must be at least 10 ms and the repetition rate $F < 100$ Hz. With the length of channels $l_{ch} = 100$ m, the repetition rate must be lower than $F < 10$ Hz.

The average intensity of the neutron flux should be possibly greater than $10^{14}$ neutrons / sec. In order to achieve this result, one requires a large pulse current of protons - tens of milliamps.

**The traditional scheme - a cylindrical cavity**

A usual linear proton accelerator with drift tubes consists of a cylindrical cavity excited by $E_{010}$ wave, which is called Alvarez structure. Drift tubes are located on the cavity axis and arranged so that there is a resonance between the wave and accelerated particle. Such a structure is resonant and calculated (at a given ratio Z / A, where Z is the particle charge, A is the atomic mass) only for one particle velocity. The spatial period $L_s$ of such a structure equals $L_s = \beta\lambda_0$, where $\beta = V / c$, V is the particle velocity, $c = 3 * 10^{10}$ cm / s is the speed of light in vacuum, $\lambda_0 = c/f_0$ is the wavelength of acceleration, $f_0$ is the frequency of the accelerating RF field. For a particle with a different speed, the drift tubes must be positioned in a different way.

In this paper, instead of a cylindrical cavity a rectangular cavity is proposed in which a few rows of drift plates are placed. In this case, plates are more convenient to use because during acceleration the beam will be displaced in a transverse direction, as in cyclotrons, and beam losses are possible on the walls of the drift tubes. When using the drift plates, the beam will move inside them in the same manner as inside the dees of a cyclotron.

The cavity must be located within two bending magnets, which are separated from each other at the distance L ≈ 15 m. Let the cavity length be l = 10 m and the field strength $E * \sin\varphi_s$ = 1 MV / m, where $\varphi_s$ is the synchronous phase, $\sin\varphi_s$ = 0.866. In this case, the proton will gain the energy $\Delta W_{cav}$ = 20 MeV per turn, and after 40 revolutions the beam can be directed to the target. The accelerator layout is shown in Fig. 1.

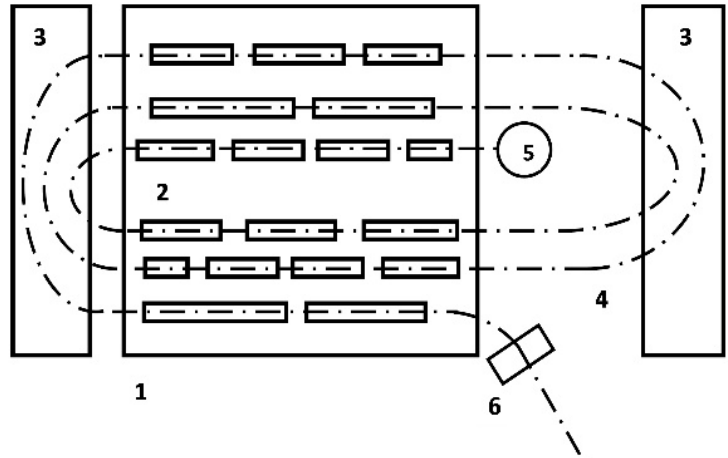

Figure 1. The structure of the accelerator. 1 - rectangular cavity with $E_{110}$ wave, 2 - drift plates fixed on bars, 3 - bending ($180^0$) magnets, 4 - beam orbits, 5 - system of external beam injection into the accelerator, 6- system of beam extraction from the accelerator.

The rectangular cavity differs from the cylindrical one, primarily, owing to several rows of drift plates installed in it, with each row of plates corresponding to the proton energy. Figure 2 shows how the drift plates can be attached to the top and bottom of the cavity.

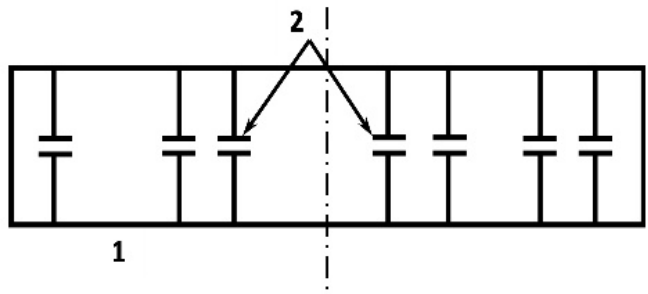

Figure 2. Cross-section of the cavity. 1 - rectangular cavity with $E_{110}$ wave, 2 – drift plates fixed on bars.

**The dimensions of the cavity**

In a magnetic field of average strength B, a proton beam rotates along the radius:

$$r_b = V / \omega_B. \qquad (1)$$

The frequency of proton rotation in a magnetic field is:

$$\omega_B = eB / (Mc\gamma). \qquad (2)$$

For the finite speed of protons $V_{fin} = 2.5 * 10^{10}$ cm / s corresponding to the proton energy $W_{fin} = 0.8$ GeV, for the average field B = 7.4 kGs, the frequency of rotation $\omega_B = 4 * 10^7$, radius of rotation $r_b = 6.25$ m. Accordingly, the diameter of the magnet pole should be approximately equal to D = 14 m. For comparison, Fig. 3 shows a general view of the PSI cyclotron, [2], where the initial energy of protons $W_{in} = 72$ MeV, final energy $W_{fin} = 590$ MeV, and diameter of the magnet is ~ 9 m.

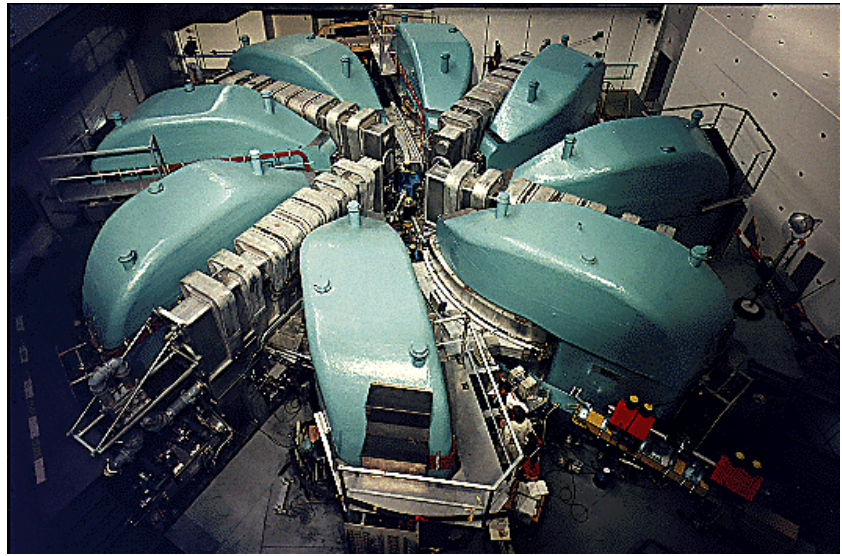

Figure 3. General view of the PSI cyclotron

The field E in the cavity is distributed vertically as $E = E_0 \sin(\pi z / a)$, where the vertical coordinate z is measured from the bottom of the cavity. At the bottom, at z = 0 the electric field is zero; at the center of the cavity, in the median plane, at z = a / 2 the field is maximum; at the top of the cavity, at z = a the field is zero again.

Along the horizontal coordinate, the field is distributed in a similar way, $E = E_0 \sin(\pi y / b)$, so that on the left vertical wall of the cavity, at $y = 0$ the field turns zero; at the center of the cavity, at $y = b/2$ the field is maximum; on the right vertical wall of the cavity, the field is again equal to zero. Thus, only the middle area of the rectangular cavity is used, which is clearly visible in the photograph of the PSI cyclotron cavity, Fig. 4, [3].

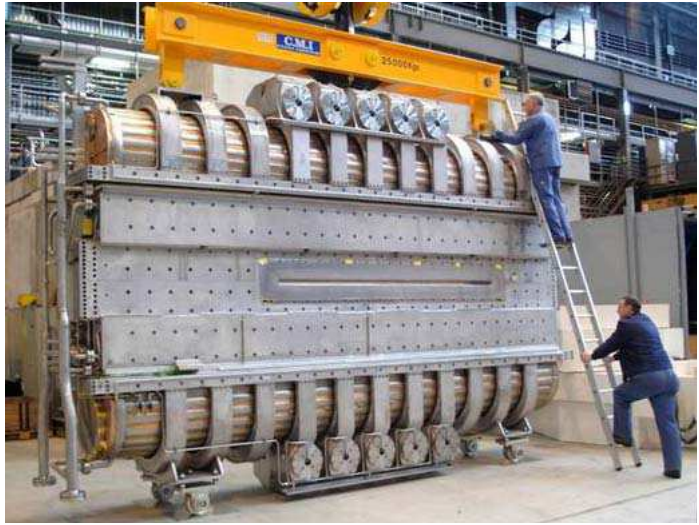

Figure 4. Accelerating cavity of the PSI cyclotron

In our case, the final radius of the beam in a magnetic field is of the order of 6.25 m, the same dimension that the "accelerating" gap must have, consequently, the transverse semi size of the cavity must be of the order $b/2 = 10$ m.

Now, let us calculate the wavelength of the cavity for the $TM_{110}$ wave or, which is the same, $E_{110}$ wave for a cavity with the transverse dimensions: $a = 1.01$ m, $b = 20$ m.

$$\lambda_0 = 2 / [(n/a)^2 + (m/b)^2]^{1/2}, \qquad (3)$$

where $n = 1$ is the vertical direction index which will be denoted as z, $m = 1$ is the index of the transverse direction (in cyclotrons it corresponds to the radial direction) which will be designated as y. Substituting the numerical values into (3), one obtains the resonance wavelength $\lambda_0 = 2.02$ m, which corresponds to the frequency $f_0 = c/\lambda_0 = 148.5$ MHz.

It is known that drift tubes do not greatly distort the natural frequency of the cavity.

Let us choose a synchronous phase for acceleration of protons in the cavity with the angle $\varphi_s = 60^0$, $\sin\varphi_s = 0.866$. The formula relating the high frequency power P to the voltage between the end walls of the cavity through which the beam passes, U, is as follows:

$$P_1 = U^2/2R_{sh}, \qquad (4)$$

where $R_{sh}$ is the so-called shunt resistance of the cavity. Shunt resistance, in its turn, can be expressed through the characteristic impedance $\rho_s$ and cavity factor Q: $R_{sh} = \rho_s * Q$. The cavity factor Q can be estimated as the ratio of the cavity volume to the skin layer volume:

$$Q = V / V_s, \qquad (5)$$

where $V = abl$ is the cavity volume, $V_s = 2\delta * (ab + al + bl)$ is the volume occupied by the skin layer $\delta$. The thickness of the skin layer $\delta = c / (2\pi\sigma\omega_0)^{1/2}$, where $\sigma$ is the conductivity, $\omega_0$ is the circular frequency. For copper, $\delta = 5.38 * 10^{-6}$ m and, substituting numerical values, one obtains: $Q = 8 * 10^4$.

In a practical system of units, the resonant frequency of the cavity can be written as:

$$\omega_0 = \pi / (\varepsilon\mu) 1/2 [1/a^2 + 1/b^2]^{1/2}, \qquad (6)$$

while the stored energy can be expressed as:

$$W = (\varepsilon / 2) \int E^2 dV. \qquad (7)$$

Since the cavity volume $V = abl$, the stored energy can be defined as:

$$W = (\varepsilon / 2) E^2 * abl * (4/\pi^2). \qquad (8)$$

Considering that $E^2 * l^2 = U^2$, where l is the longitudinal dimension of the cavity, one obtains:

$$W = (\varepsilon / 2) * U^2 * (4/\pi^2) * (ab / l). \qquad (9)$$

Substituting W into the expression for the characteristic impedance $\rho_s$, which can be written as $\rho_s = U^2/2\omega_0 W$, where $\omega_0$ is the circular frequency of the cavity $\omega_0 = 2\pi f_0$, W is the energy stored in the cavity, one finds that:

$$\rho_s = U^2/2\omega_0 W = \{2\pi / (\varepsilon\mu)^{1/2} * [1/a^2 + 1/b^2]^{1/2} (ab/l) \varepsilon * (2/\pi^2)\}^{-1}. \qquad (10)$$

Given that $(\mu/\varepsilon)^{1/2} = Z_0 = 377$ Ohm is the free space characteristic impedance, one finally obtains:

$$\rho_s = (\pi/4) * [l/(a^2 + b^2)^{1/2}] * Z_0. \qquad (11)$$

Substituting values, one can find that the characteristic impedance of the cavity $\rho_s = 148$ Ohms and, therefore, the shunt resistance of the cavity $R_{sh} = \rho_s * Q = 11.8$ MΩ. Finally, in order to obtain the voltage $U * \sin\varphi_s = 10$ MV between the ends of the cavity, one needs to introduce the power:

$$P_1 = U^2/2R_{sh} = 5.65 \text{ MW}. \qquad (12)$$

The high-frequency power required for acceleration of the beam must be greater than $P_1$, while the power transmitted to the beam at the beam current $J_b = 10$ mA and energy $W_p = 0.8$ GeV should be at least $P_2 = 8$ MW.

The GI-27A1 lamp power $P = 5$ MW, [4], so that at least two high-frequency generators are required for transmitting high-frequency power to the beam.

**Time structure of the accelerator operation**

The oscillation period of the RF field $T = 1/f_0 = 6.7$ ns. In order to provide the cavity with the working voltage, the time $\tau_{HF} = 3QT$, approximately equal to $\tau \sim 1.6$ ms, is required.

Let the average perimeter of the orbit in the accelerator be $P_{turn} = 50$ m, then the total length of proton acceleration $L_{acc} = P_{turn} * N = 2$ km, where $N = 40$ is the number of turns of the beam inside the accelerator which are required to achieve the finite energy. This number of turns is simply equal to $N = W_{fin}/\Delta W$, where $W_{fin} = 0.8$ GeV is the final energy of protons, $\Delta W_{cav} = 20$ MeV is the energy gain per revolution. The average speed of proton acceleration $V_{av}$ is approximately $c/2$, where $c = 3 * 10^5$ km/s is the speed of light in vacuum, or the average speed of protons $V_{av} = 1.5 * 10^5$ km/s. Thus, the acceleration time, i.e. the time during which the proton travels inside the accelerator, is $\tau_{acc} = L_{acc}/V_{av} \approx 15$ µs.

The power which must be transmitted to the beam $P_2 = 8$ MW is almost 2 times greater than the power needed to generate a working field in the cavity up to a value corresponding to $U * \sin\varphi_s = 10$ MV. In order to transmit power to the beam, it is required to use at least two generators. This power is to be increased gradually, in accordance with the beam's filling of the cavity. The time of enhancing the power, thus, amounts to $\tau_{acc} = 15$ μs. Then it is necessary to gradually reduce the power to avoid beam losses and electrical breakdowns. If the time duration of the "acceleration table" $\tau_{acc1} = 10$ μs, the time of acceleration $\tau_{acc2} = \tau_{acc1} + \tau_{acc} = 25$ μs.

After disconnecting the generators, the transmission power in the beam can be switched off the generator exciting the RF field in the cavity. When accelerating protons at a lower frequency, for example, f = 50 MHz, the rising time of the field $\tau_{50} = 7$ ms is comparable with the time interval between the pulses $\tau_p = 10$ ms at the repetition rate F = 100 Hz. Then it makes sense to keep up constant electric field in the cavity E * $\sin\varphi_s = 1$ MV / m and enter additional RF power only for the time of beam acceleration.

**Parameters of GI-27A1, GI-52A pulse triodes**

When working with short pulses (with a pulse at its base $\tau_b = 40$ μs), the operation of the GI-27A1 pulse triode with the repetition rate F = 10 Hz should not cause any problems. Another matter is the operation of generators that supply power for excitation of the cavity. In view of long duration of the pulse, it is also required to use several high-frequency generators as the average wattage of the GI-27A1 pulse triode is relatively small, about 25 kW [4]. The pulse duration of GI-52A can reach 2 ms at a pulse power of 5 MW and average power of 400 kW, [4]. So, it is more suitable for excitation of the cavity.

**The dynamics of protons in the bending magnets**

Let us calculate the Coulomb shift for dimensionless frequency of transverse oscillations. Now, for the initial part of the accelerator, where $\gamma \approx 1$, let us define the current of the proton beam for which the value of the Coulomb shift of the dimensionless frequency will be less than 1:

$$\Delta v_{tr} = \omega_p / \omega_{tr} < 1. \qquad (13)$$

In this formula, $\omega_p$ is the plasma oscillation frequency of protons in a cylindrical coordinate system, $\omega_p = (2\pi e^2 n_r / M)^{1/2}$, $\omega_{tr}$ is the frequency of transverse oscillations. The condition for restricting the volume density of particles in the beam can be written as:

$$\omega_p < \omega_B. \qquad (14)$$

The condition for restricting the volume density of particles in the beam (14) is written under the assumption that the frequency of transverse oscillation frequency is equal to the Coulomb frequency and the shift is equal to the frequency itself, so that $\Delta v_{tr} = 1$. Thus, in order to achieve a Coulomb frequency shift smaller than the frequency of rotation, the volume density of particles in the beam must be less than:

$$n_r < \omega_B^2 * M / (2\pi e^2). \qquad (15)$$

From (15), one finds that the peak density of the particles in the beam must be less than $n_r < 1.7 * 10^9$ protons/cm$^3$. Due to the microstructure (grouping) of the beam, the average density of protons in the beam n is by an order of magnitude lower than the peak and equals the value $n < 1.7 * 10^8$ protons/cm$^3$.

The proton current corresponding to the average particle density $n < 1.7 * 10^8$ protons/cm$^3$ must be less than:

$$J_1 < 1.6 * 10^{-19} * n * V * S, \qquad (16)$$

where $V = 4.23 * 10^9$ cm / c is the speed of the protons at the beginning of acceleration, $S = 1$ cm$^2$ is the cross-section of the beam. Substituting the values into formula (16), one obtains $J_1 < 115$ mA, which is significantly greater than the required beam current $J_b = 10$ mA.

In linear accelerators, pulsed beam currents exceeding the value $J_b = 10$ mA are achieved.

**Beam injection into the isochronous magnetic field**

Let us find the required radius of injection and beam radius after the first half-turn at horizontal beam injection. Let us assume that the mean magnetic field at the center equals $B_0 = 4$ kGs and kinetic energy of protons $W_{in} = 10$ MeV. The isochronous cyclotron frequency $\omega_B = 4 * 10^7$. According to the formula $R = V / \omega_B$, one can calculate the radius of beam rotation in the magnetic field $R_{10} = V / \omega_B = 105$ cm, where the index 10 points to the beam injection energy. This energy is gained by the proton after a single passage through the cavity.

After the second passage through the cavity, the radius of proton gyration will increase up to $R_{20} = 148$ cm, the step-orbit during the injection will amount to $\Delta r_{in} = 43$ cm. This means that the inflector (bending magnet) should have a transverse semi size of the same order.

When injecting the beam into a softly focusing magnetic field, the injector should be apparently placed vertically.

**Parallel transport of the beam**

When the beam is being injected into the magnetic field, the initial orbital radius $R_{10} = V / \omega_B = 105$ cm can be chosen in equilibrium. However, after the second passage through the cavity, the equilibrium orbit radius will increase to $R_{20} = 148$ cm, while after the passage through the bending magnet the radius remains the same, $R_{10}$, and this means that a parallel shift of the beam radius at the distance $R_{20} - R_{10} = 43$ cm is required.

Such a transfer of the beam can be accomplished by a pair of bending magnets, one of which is meant to deflect the beam outwards through the angle $\theta_{1/2} = (R_{20} - R_{10}) / L$, where $L \approx 15$ m is the distance between the bending magnets, and the other is to turn the beam through the same angle inwards. For the first half-turn the angle can be found to be $\theta_{1/2} = 26$ mrad, and with increasing energy of the beam this angle will decrease. In the proposed accelerator, the beam orbits are not parallel to the axis of the resonator.

**Extraction of particles from the accelerator**

The relative orbit step at the extraction of protons from the accelerator can be represented as:

$$\Delta r / r = \Delta V / V = \Delta \beta / \beta. \tag{17}$$

Differentiating the expression $\beta = (1-\gamma^{-2})^{-1/2}$, one obtains $\Delta \beta = 1/2 \, (1-\gamma^{-2})^{-1/2} * (2\gamma^{-3}) \Delta \gamma$ or $\Delta \beta = \Delta \gamma / \beta \gamma^3$, $\Delta \beta / \beta = \Delta \gamma / \beta^2 \gamma^3$. As a result, the orbit step of the beam in the extraction formula can be calculated:

$$\Delta r = r_{fin} * \Delta \beta / \beta = r_{fin} * \beta^{-2} \gamma^{-2} * (\Delta \gamma / \gamma). \tag{18}$$

The spatial configuration of the electric field distribution in the cavity is such that at the large radii $r \approx r_{fin}$ the energy gain per revolution is less than at the center of the accelerator. The approximate value of the energy gain is $\Delta \gamma / \gamma = 5 * 10^{-3}$, therefore, for finite values $\gamma = 1.85$ and $\beta = 0.83$ one finds that the step of the orbit radius increase in the magnetic field is of the order $\Delta r \approx 1.3$ cm.

**The possibilities of a neutron source**

Finally, let us estimate the average neutron flux for a uranium-238 target. The neutron yield per one proton for uranium is ~ 20. The beam pulse duration can be taken as $\tau_{acc2} = 25$ μs, the repetition rate F = 10 Hz. Substituting all the values into the formula, one obtains $I_n = 3 * 10^{14}$ neutrons / s.

Literature

1. http://nuclphys.sinp.msu.ru/experiment/neutr_gen/index.html
2. http://abe.web.psi.ch/accelerators/ringcyc.php
3. M. Seidel, P.A., Schmelzbach, Upgrade of the PSI Cyclotron Facility to 1.8 MW, Cyclotrons and Their Applications 2007, Eighteenth International Conference.
4. Электровакуумные приборы, Справочник, т. 16, СССР, министерство электронной промышленности, 1972,
http://www.qrz.ru/reference/tubes2/impgen.shtml